\documentstyle[aps,prl,epsf]{revtex}
\begin{document}
\twocolumn[\hsize\textwidth\columnwidth\hsize\csname@twocolumnfalse\endcsname
\title{Interlayer Quasiparticle Transport in the Vortex State of
Josephson Coupled Superconductors}
\author{
I. Vekhter$^1$, L. N. Bulaevskii$^2$, A.E. Koshelev$^3$, and  M.P. Maley$^2$ }
\address{$^1$Department of Physics, University of Guelph, Guelph, 
Ontario N1G 2W1, Canada \\
$^2$Los Alamos National Laboratory, Los Alamos, NM 87545 \\
$^3$Materials Science Division, Argonne National Laboratory, Argonne, 
Illinois 60439}
\date{\today}
\maketitle
\tighten
\begin{abstract}
We calculate the dependence of the interlayer quasiparticle conductivity,
 $\sigma_q$,
in a Josephson coupled $d$-wave superconductor on the magnetic 
field ${\bf B} \parallel c$ and the temperature $T$.
We consider a clean superconductor with resonant impurity scattering
and a dominant coherent interlayer tunneling.  When pancake
vortices in adjacent layers are weakly correlated at low $T$ the
conductivity increases sharply with $B$ before reaching an extended region
of slow linear growth, while at high $T$ it initially decreases and then
reaches the same linear regime.  For correlated pancakes $\sigma_q$
increases much more strongly  with the applied field.
\end{abstract}
\pacs{PACS numbers: 74.25.Fy, 74.50.+r, 74.72.Hs}
\vspace*{-10pt}
\twocolumn
\vskip.2pc]
\narrowtext

Experimental study of quasiparticle properties in high-temperature
superconductors so far
has focussed almost exclusively on thermodynamic and in-plane
transport properties.  Two remarkable theoretical results connect
these properties with the  symmetry of the superconducting gap:
a) a massless Dirac spectrum
of nodal quasiparticles in a d-wave superconductor
leads to
a finite density of states (DOS) at the Fermi level in presence of
impurity scattering \cite{gor} and to a universal (impurity
independent) low-temperature limit of the in-plane 
thermal conductivity, 
$\kappa_{00}\equiv\lim_{T\rightarrow 0}\kappa_{ab}(0,T)/T$,
in the absence of a magnetic field.
\cite{lee,sigmab};
b) in the vortex state the DOS is locally enhanced due to
the effect of the supercurrents around the vortex cores on the near-nodal
quasiparticles (Volovik effect) \cite{vol}.  Consequently,
the effect of the magnetic field on the
electrical and the thermal conductivities
is twofold: the net enhancement of the DOS tends to increase the
conductivity, while local changes
in the phase and amplitude of the order parameter lead to its decrease.
\cite{kubert,franz}

The enhancement of the DOS with  the magnetic field, $B$, has been invoked to
explain the increase of the specific heat with $B$ observed in
YBa$_2$Cu$_3$O$_{7-\delta}$ (YBCO).\cite{moler}
The in-plane thermal conductivity, $\kappa_{ab}$, has been
found to have a highly non-trivial dependence on the applied field
and the temperature in both YBCO and Bi$_2$Sr$_2$CaCu$_2$O$_{8+\delta}$
(Bi-2212) compounds. \cite{krish,chiao}
At low temperatures $\kappa_{ab}$
increases with $B$, while at $T\geq 5-10$ K it decreases and then often
saturates in fields above a few Tesla.  In some zero-field cooled Bi-2212
samples a hysteretic field dependence with two plateaus at different values
of $\kappa_{ab}(B)$ has been observed.\cite{krish} This indicates that
the behavior of $\kappa_{ab}(B)$ is sensitive to the vortex arrangement in
the sample.

Franz \cite{franz} has appealed to the effects
of disorder in the vortex positions to
explain the observed high-field plateau in $\kappa_{ab}(B)$.  He
has argued that randomly positioned vortices act similarly to impurities
leading to the enhancement of both the DOS and the electron scattering.
These two effects compensate each other at high fields
(in analogy to the zero-field result
being insensitive to impurity concentration)
leading to the
universal $\kappa_{ab}(B,T)/T\rightarrow \kappa_{00}$.
However, to
obtain this result Franz has made questionable approximations: he has replaced
the spatial average of the product of two Green's functions by the product
of the averages, and has assumed a
Lorentzian shape of the distribution function of the in-plane supervelocity,
which differs from a realistic
distribution in the vortex state, see below.  Clearly, this
approach cannot explain the  different (non-universal)
plateau values. \cite{krish}

Recently it has been demonstrated \cite{ju,lat} that the {\it interlayer}
quasiparticle electrical transport can be studied directly in the resistive
state of small area samples (mesas) fabricated from the Josephson coupled
superconductors such as Bi-2212.
In the energy and temperature range below $\approx
3$ meV, the experimental data are well understood in the framework of a
Fermi-liquid model for the near-nodal quasiparticles in a d-wave
superconductor assuming (i) clean limit,
(ii) resonant (unitarity) impurity scattering, and (iii)
dominant contribution  to the
$c$-axis conductivity from {\it coherent}
(conserving the in-plane momentum)
{\it interlayer tunneling}.\cite{lat}
The authors of Ref.~\onlinecite{lat}
introduced the universal quasiparticle $c$-axis conductivity
$\sigma_q(T=0,B=0)$. This quantity is insensitive to the impurity
vertex corrections which modify the in-plane conductivity as 
discussed by Durst  and Lee. \cite{sigmab}

So far the experimental data on the magnetic field dependence of interlayer
quasiparticle conductivity have been
 quite scarce. Measurements of $\sigma_c(B)$
in high fields, where $\sigma_c\approx \sigma_q$, reveal a linear increase
of $\sigma_c$ with B on the scale of 40 T. \cite{zav} For mesas Yurgens
{\it et al.} \cite{ju} have
reported a weak field dependence for $B<7$ T without
quantitative results for its functional form.

In this Letter we address the effect of vortices on the $c$-axis
quasiparticle conductivity, $\sigma_q(B,T)$, as a function of the magnetic
field ${\bf B}\parallel c$, in Josephson coupled superconductors in 
the framework of the approach used in Ref.~\onlinecite{lat}.
We find that at low (high) temperatures for $c$-axis
uncorrelated vortices $\sigma_q(B,T)$ increases (decreases) with $B$,
before reaching an extended region of slow linear-in-$B$ increase.  
In this high field regime
the increase in the DOS is largely compensated by the
enhancement of the electron scattering at tunneling, due to disorder in the
positions of vortices {\it in neighboring layers}.

To calculate the effect of vortices on the interlayer transport in the
framework of the d-wave model we employ a semiclassical approach.  \cite{vol}
The supercurrents around vortex cores lead to a Doppler shift,
$\epsilon_n({\bf k}, {\bf r})= {\bf k}\cdot{\bf v}_{sn}({\bf r})$, in the
quasiparticle spectrum, ${\cal E}({\bf k},{\bf v}_s)= E_{{\bf k}}+
\epsilon_n$.  Here ${\bf k}$ is the quasiparticle momentum  and ${\bf
v}_{sn}({\bf r})$ is the supervelocity at point ${\bf r}$ inside
layer $n$, $E_{\bf k}\approx[\xi_{\bf k}^2+\Delta^2({\bf k})]^{1/2}$ is the
quasiparticle energy at $B=0$,
$\xi_{\bf k}={\bf v}_F\cdot ({\bf k}-{\bf k}_F)$ and $\Delta({\bf
k})=\Delta_0(k_x^2-k_y^2)/k^2$.  As $c$-axis currents
are parallel to the field (in
contrast to the in-plane currents), \cite{kubert} the net interlayer
transport is determined by the spatial average of local transport
coefficients.  In the following we neglect correlations between the
positions of impurities and those of vortices, and average over the
distribution of each independently, see below.  We also neglect the
temperature dependence of $\Delta_0$ and
small effects due to Zeeman splitting.
Then the interlayer quasiparticle conductivity is
given by \cite{lat}
\begin{eqnarray}
&&\frac{\sigma_q(B,T)}{\sigma_q(0,0)}= \frac{\Delta_0}{8TN(0)}
\int_{-\infty}^{+\infty}
\frac{d\omega}{{\rm cosh}^{2}(\omega/2T)}\int d{\bf k}  \label{sigma} \\
&&\times (t^2({\bf k})/t_0^2)\left\langle
A({\bf k},\omega+\epsilon_n)A({\bf k},\omega+\epsilon_{n+1})\right\rangle,
\nonumber
\end{eqnarray}
where $\sigma_q(0,0)=2e^2t_0^2N(0)s\eta/\pi\hbar\Delta_0$ is the
universal $c$-axis conductivity, $N(0)$ is the 2D DOS, $t({\bf k})$ is the
interlayer transfer integral, $t_0=t({\bf k}_g)$, ${\bf k}_g$ are the 
positions 
of the gap nodes on the Fermi surface, and $0<\eta<1$ is the weight for the
coherent tunneling.
Further, $\langle ...\rangle$ denotes the spatial average, and
$$
A({\bf k},\omega)\approx
\left(1+\frac{\xi_{\bf k}}{E_{\bf k}}\right)\frac{L(\omega, E_{\bf k})}{2}+
\left(1-\frac{\xi_{\bf k}}{E_{\bf k}}\right)\frac{L(\omega,-E_{\bf k})}{2}
$$
is the spectral density
averaged over impurities with
$L(\omega,E)=\gamma(\omega)/\pi[(E-\Omega(\omega))^2+\gamma^2(\omega)]$.
The functions $\omega-\Omega(\omega)$ and $-\gamma(\omega)$ are the real and
the imaginary part of the self-energy respectively.\cite{lee}
In the unitarity limit
the effective scattering rate of quasiparticles
is $\gamma(\omega)\approx
\gamma_0-\omega^2/8\gamma_0$ when $\omega\ll\gamma_0$ and
$\gamma(\omega)\approx \pi \gamma_0^2/2 |\omega |$ when
$\omega\gg\gamma_0$, where
$\gamma_0 \approx (\hbar\nu_0\Delta_0)^{1/2}$ and $\nu_0$ is the bare
scattering rate.   The renormalized frequency is $\Omega(\omega)\approx
\omega/2$ when $\omega\ll\gamma_0$ and $\Omega(\omega)\approx \omega$ if
$\omega\gg\gamma_0$.

At low temperatures $T\ll\Delta_0$ the quasiparticle current comes mainly
from the regions near the gap nodes. 
In the vicinity of a node we can linearize the quasiparticle
spectrum
and obtain from Eq.~(\ref{sigma})
\begin{eqnarray}
&&\frac{\sigma_q(B,T)}{\sigma_q(0,0)}=\int_{-\infty}^{+\infty}d\omega
\int_0^{\infty}dE\frac{3\pi^2E}
{8T{\rm cosh}^2(\omega/2T)}\times \label{sigma1} \\
&&\langle L(\omega+\epsilon_{n+1},E)[L(\omega+\epsilon_{n},E)
+(1/3)L(\omega+\epsilon_{n},-E)]\rangle,
\nonumber
\end{eqnarray}
where $\epsilon_n({\bf r})={\bf k}_g\cdot {\bf v}_{sn}({\bf r})$ and 
we have set $t({\bf k})\simeq t_0$.
The characteristic energy scale for the
Doppler shift, $\epsilon_n$,
is $\epsilon_B=\hbar v_F/a$, where
$a=(\Phi_0/B)^{1/2}$ is typical
intervortex distance. The semiclassical approach
is valid for $\epsilon_B\ll\Delta_0$, i.e. for fields $B\ll B_{\Delta}$,
where $B_{\Delta}=\Phi_0\Delta_0^2/\hbar^2v_F^2$.
This approach also does not account correctly for the quasiparticles in the
regions near the vortex cores, which
leads to corrections of the order of
$\delta\sigma_q/\sigma_q\approx \epsilon_B^2/\Delta_0^2=B/B_{\Delta}$.
Another important field scale can be obtained by comparing $\epsilon_B$ with
the scattering rate $\gamma_{0}$; the field
$B_{\gamma}=\Phi_0\gamma_0^2/\hbar^2 v_F^2$ separates the regimes of
impurity dominated and field dominated behavior.
Using the parameters $\Delta_{0}\approx 25$ meV,
$\gamma_0\approx 2-3$ meV\cite{lat} and $v_F\approx 1.5-2.5\cdot 10^7$ cm/s
\cite{ding} we obtain $B_{\gamma}\approx 0.2-0.6$ T and $B_{\Delta}\approx
40-80$ T.

We now rewrite the spatial average in Eq.(\ref{sigma1}) as the 
average over the probability distribution of the Doppler shift, $\epsilon$, 
which is fully determined by the vortex arrangement.  In Bi-2212 crystals 
the  3D vortex lattice is destroyed by pinning at the 
``second peak field'' $B\approx 0.02$-$0.05$ T (see, e.g., \cite{SecPeak}),  
and at higher fields pancakes in neighboring layers are only weakly correlated.
Consequently, we consider the limits of
$c$-axis-correlated and
uncorrelated pancakes.  In both cases the average
depends only on the probability
distribution function of
$\epsilon$ in a single layer, ${\cal P}(\epsilon)$,
which is related to the distribution
$P(p_{x})$ of a single component of supermomentum $p_{x}=2 m v_{sx}$,
where $m$ is the effective mass,
by ${\cal P}(\epsilon)=(2/v_{F})P(p_{x}=2\epsilon/v_{F})$.
The function
$P(p_{x})$ is determined by the pancake configuration
$\{{\bf R}_{i}=(X_i, Y_i)\}$, and,
when $a$ is smaller than the London penetration depth $\lambda_{ab}$,
is given by $P(p_{x})=\langle
\delta(p_{x}-\hbar \sum_{i}Y_{i}/R_{i}^{2})\rangle_{\{{R_{i}}\}}$.
For an isolated
vortex $p\propto\hbar/R $ where $\xi_{ab}\ll R\ll a$ is the distance
from the pancake
center, and
$\xi_{ab}$ is the coherence length.  Consequently,
${\cal P}(\epsilon)$ has a
universal tail
${\cal P}(\epsilon) =\pi\epsilon_{B}^{2}/(8\epsilon^{3})$ at
$\epsilon_B \ll \epsilon \ll \Delta_0$,
while its  behavior at $\epsilon\lesssim\epsilon_B$
depends on the actual positions of vortices.
However, in absence of correlation between the positions
of vortices and impurities ${\cal P}(\epsilon)$ involves a single
energy scale $\epsilon_B$ for {\it any} $\epsilon$,
and depends on $\epsilon$ only via
$\epsilon/\epsilon_B$, i.e.
${\cal P}(\epsilon) d\epsilon={\cal P}_B(\epsilon/\epsilon_B)
d\epsilon/\epsilon_B$.
In the clean limit
we can extend the asymptotic behavior ${\cal
P}(\epsilon)\propto 1/\epsilon^3$ to infinity when the integral over
$\epsilon$ in Eq.~(\ref{sigma1}) converges.  Therefore, importantly, up to
terms of the order of $\gamma_{0}/\Delta_0$, the
conductivity depends only on the dimensionless parameters
$\epsilon_B/\gamma_0\equiv \sqrt{B/B_{\gamma}}$ and $T/\gamma_0$.

For c-axis-correlated vortices $\epsilon_n\approx
\epsilon_{n+1}$, and the average in Eq.~(\ref{sigma1}) means
$\langle {\cal F}(\epsilon_{n},\epsilon_{n+1}) \rangle \approx \int
d\epsilon {\cal P}(\epsilon){\cal F} (\epsilon,\epsilon)$.
At low temperatures, $T\ll\gamma_0$, both in weak ($B\ll B_\gamma$) and
in strong ($B\gg B_\gamma$) fields the leading field-dependent part of
the conductivity varies as $(\epsilon/\gamma_0)^2$. 
We therefore separate $\sigma_q$ into the contributions from the regions 
$\epsilon<\gamma_0$ and $\epsilon>\gamma_0$, 
and use the asymptotic behavior of $\Omega(\omega)$ and $\gamma(\omega)$
in each case to estimate
\begin{equation}
\frac{\sigma_q(B,0)-\sigma_q(0,0)}{\sigma_q(0,0)}\approx
\frac{1}{6}\frac{\langle\epsilon^2\rangle_{\epsilon<\gamma_0}}
{\gamma_0^2}+\frac{3\pi^2}{8}
\frac{\langle\epsilon^2\rangle_{\epsilon>\gamma_0}}{\gamma_0^2}.
\end{equation}
%
%
As the average is dominated by the tail 
of ${\cal P}(\epsilon)$ we can evaluate $\sigma_q$ in weak and 
strong fields with logarithmic accuracy. Moreover, we interpolate between 
the two limits to obtain also a qualitative description at $B\sim B_\gamma$,
\begin{eqnarray}
\nonumber
&&\sigma_q(B,T)/\sigma_q(0,0)\approx 1+\pi^2T^2/18\gamma_0^2 \\
&&+\frac{3\pi^3 B}{64 B_{\gamma}}\left
[\ln\frac{B_\Delta}{(B^2+B_{\gamma}^2)^{1/2}}+\frac{4}{9\pi^2}\ln\frac{
(B^2+B_{\gamma}^2)^{1/2}}{B}\right].
\label{cryst}
\end{eqnarray}
Therefore at low $T$ for the 3D-ordered case 
 $\sigma_q$ increases quasi-linearly with $B$
in the entire field range up to $B_{\Delta}$.

For the $c$-axis uncorrelated vortices the average in Eq.~(\ref{sigma1})
has to be taken independently in each layer,
$\langle {\cal F}(\epsilon_n,\epsilon_{n+1}) \rangle =\int
d\epsilon_1 {\cal P}(\epsilon_1)\int
d\epsilon_2 {\cal P}(\epsilon_2){\cal F} (\epsilon_1,\epsilon_2)$.
Again, for $T\ll \gamma_0$ and  $B\ll B_{\gamma}$ we expand
in $\epsilon_n/\gamma_0$, and obtain 
\begin{equation}
\frac{\sigma_q(B,T)}{\sigma_q(0,0)}\approx 1+\frac{\pi^2T^2}{18\gamma_0^2}
+\frac{\pi B}{96B_{\gamma}}\ln\frac{B_{\gamma}}{B}.
\end{equation} 
At high fields the variation of $\epsilon_n$ is of the order of
$\epsilon_B\gg\gamma_0$ and consequently $\langle
L(\omega+\epsilon_n,E)\rangle\approx {\cal P}(E-\omega)$.
Then for $B_{\gamma}\ll B\ll B_\Delta$ the
conductivity is
given by
\begin{equation}
{\sigma_q (B, T)\over \sigma_q(0,0)}=C_1+{B\over B_0}, \ \
C_1=2\pi^2\int_0^{\infty}d\epsilon \epsilon {\cal P}^2(\epsilon),
\label{hf}
\end{equation}
The most important contribution to the linear, in $B$, term
is due to the increase in the tunneling away from the nodes,
$t({\bf k})\simeq t_0+ t_1\phi^2$, where $\phi$ is the angle between {\bf k}
and the nodal direction, and $t_1\gg t_0$,\cite{ioffe}
which yields $B_0\sim (t_0/t_1) B_\Delta$. Other contributions,
due to deviations of the
quasiparticle spectrum from the massless Dirac form of the linearized
dispersion and due to corrections to
the semiclassical approximation in the vicinity of the vortex cores,
only enhance $\sigma_q$ on the scale of $B_\Delta \gg B_0$.
Very importantly, due to scaling of
${\cal P}(\epsilon)={\cal P}_B(\epsilon/\epsilon_B)/\epsilon_B$ (see above),
$C_{1}$ in
Eq.~(\ref{hf}) is a constant which
depends solely on the {\it shape} of the distribution
${\cal P}(\epsilon)$, but not on the magnetic field.
If ${\cal P}_B(x)$ is monotonous   $C_1$ depends on 
its asymptotic decay.  For the Lorentzian
(when vortices act {\it exactly} as the
impurity scatterers do at $B=0$) ${\cal P}_B(x)\propto x^{-2}$ at $x\gg 1$,
and $C_1=C_L=1$,  while for the Gaussian distribution
$C_1=C_G=\pi/2$.  In the vortex state ${\cal P}_B(x)\propto x^{-3}$
at large $x$,
and we expect  $1<C_1<\pi/2$.  Therefore to find
$\sigma_q$ we need accurate information on pancake arrangement 
to determine the supervelocity distribution
${\cal P}(\epsilon)$.

Below the irreversibility line $T_{irr}(B)$ at high fields and in the
vortex liquid state pancakes do not possess long range order inside layers
due to pinning and thermal fluctuations.  To calculate $\sigma_q(B,T)$
in these regimes we use the distribution function $P(p_{x})$ obtained by
numerical simulation of the 2D pancake liquid at different values of the
dimensionless
temperature $t=T_{eff}/2\pi E_0$, which characterizes pancake disorder
inside the layers.  Here $E_0=\Phi_0^2s/16\pi^3\lambda_{ab}^2$ is the
characteristic energy of pancake interaction. In the
liquid state $T_{eff}=T$, while  in the glass state it is reasonable to
take $T_{eff}\approx T_{irr}(B)$.  The calculated function
$P(p_{x})$ is shown in Fig.~1 for several values of  $t$;
the inset shows that the parameter
$C_1$ grows with $t$ from 1.08 to 1.22.
The full field dependence of $\sigma_q(B,T)/\sigma_q(0,0)$ at $T\ll\gamma_0$
obtained from these distribution functions is shown in Fig.~2.
In computing $\sigma_q$ we have determined
$\gamma(\omega)$ and $\Omega(\omega)$ self-consistently \cite{note}
in the unitarity limit, choosing the scattering rate so that
$\gamma_{0}= 0.1 \Delta_{0}$, typical of Bi-2212 samples.\cite{lat}

Due to short range correlations inherent to strongly interacting vortices
at $B>\Phi_0/\lambda_{ab}^2$ the calculated function $P(p_x)$
differs significantly from the Gaussian (obtained in Ref.~\onlinecite{yu}
assuming uncorrelated and random vortex positions).
The distribution function depends on a single variable
$\epsilon/\epsilon_B$ only if the positions of vortices and those of
impurities are uncorrelated; this is the case for a vortex solid and
for the liquid phase in presence of weak pinning.  In general, below the
irreversibility line impurities act as pinning centers, and the
distribution function depends on two variables, $\epsilon/\epsilon_B$ and
$B/\Phi_0 n_i$, where $n_i$ is the impurity concentration in a layer.
We believe nevertheless that in that regime our approach gives at least the
correct qualitative behavior.

At $T\gg \gamma_0$, we also consider
the intermediate fields,
$B_{\gamma}\ll B\ll B_T=\Phi_0T^2/\hbar^2v_F^2$, when
the quasiparticle concentration is determined  by the temperature
and the primary effect of vortices is to increase the scattering at 
tunneling. Consequently $\sigma_q$ decreases with field
\begin{equation}
\frac{\sigma_q(B,T)}{\sigma_q(0,0)}\approx
C_2\sqrt{\frac{B_T}{B}},
\ \
C_2=\frac{3}{2}\pi^2\ln 2\int_0^{\infty}
dx{\cal P}_B^2(x).
\end{equation}
At $B\gg B_T$ Eq.~(\ref{hf}) holds.  Therefore, for $c$-axis uncorrelated
pancakes at $T\ll\gamma_0$,
$\sigma_q(B,T)$ increases quasi-linearly with $B$ with the slope
$1/B_\gamma$ when   $B \ll B_{\gamma}$,
and  with a smaller slope $\sim 1/B_0$, for
$B_\gamma\ll B\ll B_\Delta$.
At higher temperatures, $T\geq \gamma_0$,
$\sigma_q(B,T)$ {\it decreases}  with field for $B\lesssim B_T$, before
crossing over to the slow linear growth with the slope $\sim 1/B_0$,
see Fig.~3. Here the conductivity has been computed
with a temperature-independent $\Delta_0$. In fact
$\Delta_0$ is reduced with increasing $T$, and the plateau values increase.
Note that 
linear interpolation of $\sigma(B,T)$ from high fields does {\it not}
yield $\sigma_q (0, T)$.

To conclude, we find  that:
a) the field dependence of the quasiparticle interlayer
conductivity is sensitive to the structure of vortex state; b) for the
$c$-axis uncorrelated vortices the field dependence of the interlayer
conductivity at low and high temperatures is quite different at low fields
but becomes similar in high fields $B\gg \max[B_T, B_\gamma]$.

We thank M. J. Graf for helpful discussions.
This work was supported by the Los Alamos National Laboratory under the
auspices of the U.S. Department of Energy.  Work in Argonne was supported
by the NSF Office of the Science and Technology Center under contract
No.~DMR-91-20000 and by the U.S. DOE, BES-Materials Sciences, under
contract No.~W-31-109-ENG-38. I.V. acknowledges the hospitality of Centre
\'Emile Borel and Aspen Center for Physics, where part of this work was
done.


\begin{figure}
\epsfxsize=3.2in \epsffile{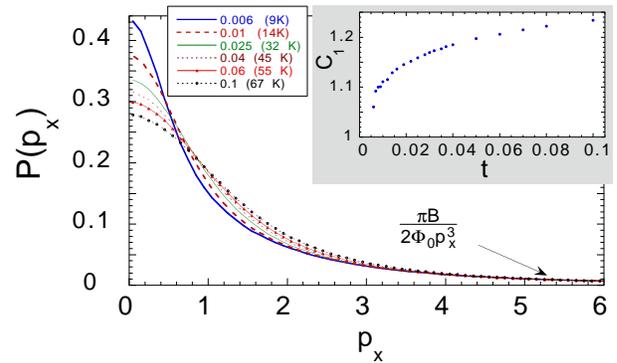}
\caption{The distribution
functions $P(p_x)$ of the $x$-component of supermomentum. Here $p_x$ is
measured in units $\hbar/a_0$, where $a_0=(2\Phi_0/\protect\sqrt{3}B)^{1/2}$ is
the lattice constant of a triangular vortex lattice.  $P(p_x)$
has been calculated using Langevin dynamics simulations of a
2D liquid at different reduced temperatures $t=T/2\pi E_0$.
The corresponding absolute temperatures obtained assuming
$\lambda_{ab}=200{\rm nm}(1-T^2/T_c^2)^{-1/2}$ are given in brackets.
The distribution function of the Doppler shifts ${\cal P}(\epsilon)$
is related to $P(p_x)$ by ${\cal P}(\epsilon)=
(2a_0/\hbar v_F)P(p_x=2a_0\epsilon /\hbar v_F)$.
Inset: temperature dependence of the parameter $C_1$ in
Eq.~(6).}
\label{Fig-Dist}
\end{figure}

\begin{figure}
\epsfxsize=3.2in \epsffile{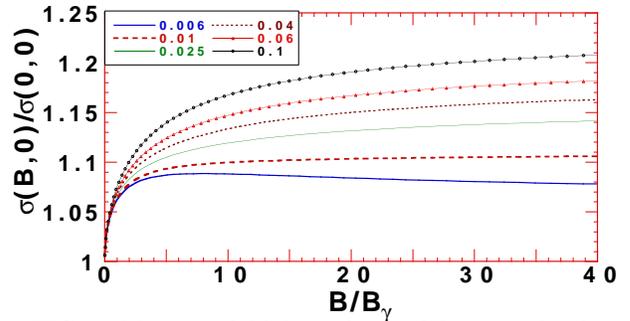}
\caption{Magnetic field dependence of the normalized
quasiparticle conductivity $\sigma_q(B,0)/\sigma_q(0,0)$ at
low temperatures $T\ll \gamma_0$ for $c$-axis uncorrelated vortex state
calculated using functions $P(p_x)$ shown
Fig.~1. Here we do not show
the weak linear field dependence at $B/B_\gamma\gg 1$, see
Eq.~(6).}
\label{Fig-sigB}
\end{figure}

\begin{figure}
\epsfxsize=3.2in \epsffile{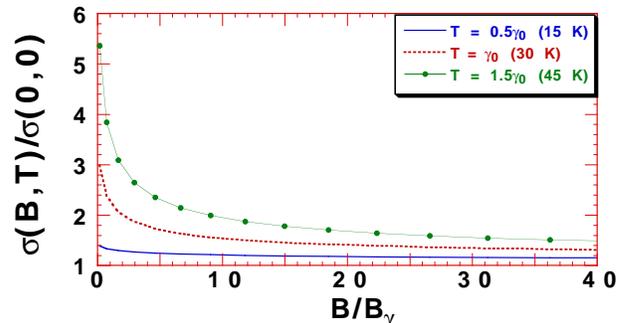}
 \caption{Field dependence of the quasiparticle conductivity at
temperatures $T\ge 0.5\gamma_0$, computed for the pancake liquid state with
$P(p_x)$ obtained by simulations (Fig.~1) and
$\gamma_0=0.16 E_0\approx 30$ K.}
\label{Fig-sigBT}
\end{figure}


\begin{thebibliography}{100}
\bibitem{sigmab} Ref.~\onlinecite{lee} has also predicted the
universal limit for the in-plane electrical conductivity. However, recent
work has shown that the universality is violated if vertex and Fermi-liquid
corrections are accounted for, see A.~C.~Durst and P.~A.~Lee, cond-mat/9908182.
\bibitem{note} We ignore here the change of the internode scattering
with the Doppler shift. \cite{kubert}
If included, it only affects the intermediate field
regime, and does not change the low and high field limits.
\end{thebibliography}
\end{document}